# THE HIDDEN SCALAR LAGRANGIANS
# WITHIN HORNDESKI THEORY

by

Gregory W. Horndeski
2814 Calle Dulcinea
Santa Fe, NM 87505-6425
email:
horndeskimath@gmail.com

March 30, 2020

## ABSTRACT

In this essay I show that there exists a new way to obtain scalar-tensor field theories by combining a special scalar field on the cotangent bundle with a scalar field on spacetime. These two scalar fields act as a generating function for the metric tensor. When using these two scalar fields in the Horndeski Lagrangians, we discover, while seeking Friedmann-Lemaître-Robertson-Walker type cosmological solutions, that hidden in the Horndeski Lagrangians are non-degenerate second-order scalar Lagrangians. In accordance with Ostrogradsky's work, these hidden scalar Lagrangians lead to multiple vacuum solutions, and thereby predict the existence of the multiverse. The multiverse is comprised of numerous different types of individual universes. *E.g.,* some begin explosively, and then coast along exponentially forever at an accelerated rate, while others begin in that manner, and then stop expanding and contract.

**Section 1: Introduction**

In traditional scalar-tensor field theories of gravitation the field variables are the components, $g_{ij}$, of a Lorentzian metric tensor g on a 4-dimensional manifold M, along with a scalar field φ on M. If we require the field equations to be derivable from a Lagrange scalar density, and be at most of second-order in the derivatives of the field variables, then the most general field equations of that type can be obtained from (*see,* Horndeski [1], [2], Deffayet, *et al.,*[3], Kobayashi, *et al.*, [4])

$$L_H := L_2 + L_3 + L_4 + L_5 \qquad \text{Eq.1.1}$$

where

$$L_2 := g^{½}K(\varphi,X) \qquad \text{Eq.1.2}$$

$$L_3 := g^{½}G_3(\varphi,X)\Box\varphi \qquad \text{Eq.1.3}$$

$$L_4 := g^{½}G_4(\varphi,X)R - 2g^{½}G_{4,X}(\varphi,X)[(\Box\varphi)^2 - \varphi^{ab}\varphi_{ab}] \qquad \text{Eq.1.4}$$

and

$$L_5 := g^{½}G_5(\varphi,X)\varphi_{ab}G^{ab} + ⅓g^{½}G_{5,X}(\varphi,X)[(\Box\varphi)^3 - 3\Box\varphi(\varphi^{ab}\varphi_{ab}) + 2\varphi^a{}_b\varphi^{bc}\varphi_{ca}], \quad \text{Eq.1.5}$$

where $g:= |\det(g_{ab})|$, $X:=g^{ab}\varphi_{,a}\varphi_{,b}$, ",X" $:= \partial/\partial X$, and $\varphi_{ab}$ is the second covariant derivative of φ. Definitions of other quantities can be found in [2].

One way of working with Horndeski scalar-tensor theory is to choose a form of the coefficient functions K, $G_3$, $G_4$, $G_5$, and then compute the associated Euler-Lagrange equations. After doing that solutions are sought to those equations, which usually involves making certain symmetry assumptions about the spacetime of interest. For cosmological purposes we can choose an ansatz metric of the FLRW (:=

Friedmann-Lemaître-Robertson-Walker) form, which has line element

$$ds^2 = -dt^2 + a(t)^2[\,(1-kr^2)^{-1}dr^2 + r^2d\theta^2 + r^2\sin^2\theta d\varphi^2] \qquad \text{Eq.1.6}$$

where k is a constant with units (length)$^{-2}$, r has units of length, a(t), θ and $\varphi$ are unitless, and the speed of light, c=1. k determines the curvature of the spaces t=constant, which are 3-surfaces of constant curvature. To make matters even simpler, I shall choose k=0 in Eq.1.6, and so our line element becomes

$$ds^2 = -dt^2 + a(t)^2[du^2 + dv^2 + dw^2], \qquad \text{Eq.1.7}$$

where (t,u,v,w) are the standard coordinates of $\mathbb{R}^4$, and the spaces t=constant are flat. One could use the metric given in Eq.1.7, along with $\varphi=\varphi(t)$, in the Euler-Lagrange equations obtained from a suitable $L_H$ by varying $g_{ij}$ and $\varphi$, to obtain cosmological field equations. However, there is another way of obtaining cosmological field equations from $L_H$, which I shall now describe.

**Section 2: Lorentzian Cofinsler Spaces and the Hidden Lagrangians**

In [5] I show that it is possible to construct a geometry on the cotangent bundle, T*M, of a spacetime M, which is similar to Finsler Geometries on TM. I call this new geometry Lorentzian Cofinsler Geometry. This geometry is based on a function f, defined on an open submanifold N of T*M, where πN = M, and π: T*M→M is the natural projection. If x is a chart of M with domain U we let (χ,y) denote the corresponding standard chart of T*M with domain $\pi^{-1}U$, where χ:=x∘π and y: $\pi^{-1}U\to\mathbb{R}^4$ is defined by $y(\omega_i dx^i):=(\omega_1,\omega_2,\omega_3,\omega_4)$, so $y_j(\omega_i dx^i) = \omega_j$. In order for f to define a Lorentzian Cofinsler Geometry we require that $\forall\ \omega\in\pi^{-1}U\cap N$ the matrix

$$[g^{ij}(\omega)]:=\left[\frac{\frac{1}{2}\partial^2 f(\omega)}{\partial y_i \partial y_j}\right] \qquad \text{Eq.2.1}$$

defines a Lorentzian quadratic form on $\mathbb{R}^4$. In [5] it is shown that this restriction on f is independent of the chart x chosen for M. The triple (M,N,f) is called a Lorentzian Cofinsler Space (=:LCS), denoted by $CF^4$, and the scalar field f is called a Lorentzian Cofinsler Function (=:LCF).

If $V_4 = (M,g)$ is a Lorentzian spacetime then it naturally gives rise to a LCS when f is locally defined on T*M by $f:= g^{ij}\circ\pi\ y_i y_j$. Such a LCS is said to be trivial.

If $CF^4 = (M,N,f)$ is a LCS and $\varphi$ is a scalar field on M $\ni d\varphi\subset N$, then f and $\varphi$ naturally give rise to a Lorentzian metric tensor, $g_\varphi$, on M whose contravariant components are locally defined by

$$[g_\varphi^{ij}] := \left[\frac{\frac{1}{2}\partial^2 f(d\varphi)}{\partial y_i \partial y_j}\right]. \qquad \text{Eq.2.2}$$

I shall refer to a scalar-tensor field theory based upon a Lorentzian Spacetime $V_4$ = (M,g), and scalar field $\varphi$, for which g is obtained from a LCS by means of Eq.2.2, as a scalar-scalar theory. In [5] I discuss the problem of trying to obtain field equations for a scalar-scalar theory using a Lagrangian, L, of a scalar-tensor field theory. One suggested method of doing so is to choose a value of f suitable for the problem at hand, and then putting f and $\varphi$ into L, to obtain a Lagrangian that is a function of $\varphi$ and its derivatives, along with the local coordinates of M. If f is chosen to only depend upon the $y_i$ coordinates, then L would locally have the form

$$L = L(\varphi; \varphi_{,a}; \varphi_{,ab}; \ldots).$$

This Lagrangian will be referred to as a hidden scalar Lagrangian.

Of particular interest to me are the hidden scalar Lagrangians within $L_H$, when f is chosen to give us a FLRW metric of the form given in Eq.1.7. To that end we choose

$$f := -y_t^2 + y_t^{-2}((y_u)^2 + (y_v)^2 + (y_w)^2).  \qquad \text{Eq.2.3}$$

In [5] it is shown that f defines a LCF in $T^*\mathbb{R}^4$. If $\varphi=\varphi(t)$, then the metric g on $\mathbb{R}^4$ obtained using Eq.2.2 is given by (I have dropped the $\varphi$ on g)

$$ds^2 = -dt^2 + (\varphi')^2(du^2 + dv^2 + dw^2), \qquad \text{Eq.2.4}$$

where $\varphi' := \frac{d\varphi}{dt}$, and we assume that $\varphi' > 0$. In terms of geometrized units $\varphi'$ is unitless and plays the roll of a scale factor.

To illustrate the hidden Lagrangian approach described above using the metric presented in Eq.2.4 we need to choose a suitable $L_H$. Now recently it has been shown that gravitational waves must travel at c (Abbott, *et al.,*[6]). This implies that $G_5 \approx 0$, and $G_4 = G_4(\varphi)$ in $L_H$ to generate a scalar-tensor theory compatible with observations (*see,* Baker*, et al.,* [7], Creminelli & Vernizzi [8], Sakstein & Jain [9], Ezquiaga &

Zumalaćarregui [10]). Thus let us consider the Lagrangian $\mathcal{L} = L_2 + L_4$. To further simplify our task, assume that

$$K := A_2 \varphi^\mu |X|^\nu = A_2 \varphi^\mu (\varphi')^{2\nu} \text{ and } G_4 := A_4 \varphi^\eta \qquad \text{Eq.2.5}$$

where $\mu, \nu, \eta, A_2$ and $A_4$ are numbers to be determined. When Eqs.1.2, 1.4. 2.4 and 2.5 are combined we get a Lagrangian which is third order in $\varphi$. However, the third order term can be incorporated into a divergence. When this divergence is dropped the resulting Lagrangian is

$$\mathcal{L}_f = A_2 \varphi^\mu (\varphi')^{2\nu+3} - 6 A_4 \eta \varphi^{\eta-1} (\varphi')^3 \varphi'' - 6 A_4 \varphi^\eta \varphi' (\varphi'')^2. \qquad \text{Eq.2.6}$$

At this point savvy readers have probably leaped to their feet screaming that $\mathcal{L}$ is a non-degenerate second-order Lagrangian, rotten with Ostrogradsky [11] type singularities (*see,* Woodard [12]), and should be dismissed with immediately. I say, let's not be too hasty. One of the problems associated with a Lagrangian like $\mathcal{L}$ is that it can lead to multiple vacuum solutions. This might not be a bad thing if we are looking for an equation whose solutions predict the existence of a multiverse, since a multiverse would need separate vacuums for each of the individual universes. That is precisely what $\mathcal{L}$ predicts.

**Section 3: Cosmological Solutions**

Our observations of the U (:=Universe) suggest that it began explosively, and then the rapid expansion subsided giving rise to a state of gradual accelerated expansion, whose duration is presently unknown. So let $\mathcal{H}$ denote the Hamiltonian of $\mathcal{L}$. We seek solutions to $\mathcal{H}=0$ (*i.e.,* vacuum solutions), of the form $\varphi=\alpha e^{\beta t}$ and $\varphi=\gamma(k_1 t+k_2)^q$, where $\alpha$, $\beta$, $\gamma$, $k_1$, $k_2$ and $q$ are real constants, with appropriate units. In [5] I show that for various choices of $\mu$, $\nu$, $\eta$, $A_2$ and $A_4$ we can obtain numerous solutions for $\varphi$ of the required form. Solutions with $1<q<2$ exist, which (when $k_2=0$) correspond to singular metrics as $t \to 0^+$, while we can't get exponential solutions that do that.

For example, when we choose $\eta=-5/3$ in $\mathcal{L}$, we must have $\nu=1$, $\mu=-11/3$, $q=3/2$ and $A_2=-14A_4/3$ (with $A_4$ arbitrary), to have $\varphi$ solutions of the required form. For these solutions $\alpha$, $\beta$, $\gamma$, $k_1$ and $k_2$ are arbitrary, up to the demand that $\varphi'>0$. In the solutions $\varphi=\gamma(k_1 t+k_2)^{3/2}$, the scalar curvature $R=0$, while $R=12\beta^2$ when $\varphi=\alpha e^{\beta t}$.

Say we start out at $t=0^+$ in a solution of the form $\varphi_1=\gamma(k_1 t)^{3/2}$, with $\gamma k_1>0$, to assure that $\varphi_1'>0$. As our U expands, originally with infinite velocity, at some time $t_1$ we encounter an exponential solution $\varphi_2=\alpha e^{\beta t}$, which is such that $\varphi_1'(t_1)=\varphi_2'(t_1)$. At that time we can then jump from the $\varphi_1$ solution to the $\varphi_2$ solution, which is an accelerating solution, when $\beta>0$. This jump would lead to a discontinuity in $\varphi$, but the metric defined by Eq.2.4 would remain continuous. At a latter time we could then jump to another decreasing solution of the form $\varphi=\alpha' e^{\beta' t}$ or $\varphi=\gamma'(k_1' t+k_2')^{3/2}$ keeping the metric continuous.

Hence this hidden Lagrangian approach permits us to construct a multiverse populated by U's that begin catastrophically, then either coast along accelerating gradually forever, or which stop accelerating, and then return to φ'=0.

**Acknowledgemnt**

I would like to thank Dr. Kazufumi Takahashi for valuable discussions on some of the topics dealt with in this essay.